\documentclass[runningheads]{llncs}
\usepackage{url}
\makeatletter
\AtBeginDocument{%
  \def\doi#1{\url{https://doi.org/#1}}}
\makeatother

\usepackage{multirow}
\usepackage{svg}
\usepackage{array}
\usepackage{graphicx}
\usepackage{subcaption}

\usepackage{xcolor}
\usepackage{soul}  
\usepackage{todonotes} 
\begin{document}
\title{
Distributed Simulation for Digital Twins of Large-Scale Real-World DiffServ-Based Networks}

%
%
\author{Zhuoyao Huang\textsuperscript{\dag}\inst{1,2} \and
Nan Zhang\textsuperscript{\dag}\inst{4}\orcidID{0000-0002-5728-0440} \and
Jingran Shen\inst{4}\orcidID{0000-0002-1696-9663} \and
Georgios Diamantopoulos\inst{4,6}\orcidID{0009-0007-6153-3141} \and
Zhengchang Hua\inst{4,7} \and
Nikos Tziritas\inst{5} \and
Georgios Theodoropoulos\textsuperscript{*}\inst{3,4}
}

\authorrunning{Z. Huang, N. Zhang et al.}
\titlerunning{Distributed Simulation for Digital Twins of Networks}
%
\institute{State Key Laboratory of Mobile Network and Mobile Multimedia Technology, Shenzhen, China \and
Shenzhen ZTE Communication Technology Service Co., Ltd, Shenzhen, China \and
Research Institute of Trustworthy Autonomous Systems (RITAS), Shenzhen, China \and
Southern University of Science and Technology (SUSTech), Shenzhen, China \and
University of Thessaly, Lamia, Greece \and
University of Birmingham, Birmingham, UK \and
University of Leeds, Leeds, UK
}
\maketitle              

\def\thefootnote{\dag}\footnotetext{Co-first author.}
\def\thefootnote{\arabic{footnote}}
\def\thefootnote{*}\footnotetext{Corresponding author.}
\def\thefootnote{\arabic{footnote}}

\begin{abstract}
Digital Twin technology facilitates the monitoring and online analysis of large-scale communication networks. Faster predictions of network performance thus become imperative, especially for analysing Quality of Service (QoS) parameters in large-scale city networks.
Discrete Event Simulation (DES) is a standard network analysis technology, and can be further optimised with parallel and distributed execution for speedup, referred to as Parallel Discrete Event Simulation (PDES).
However, modelling detailed QoS mechanisms such as DiffServ requires complex event handling for each network router, which can involve excessive simulation events.
In addition, current PDES for network analysis mostly adopts \textit{conservative} scheduling, which suffers from excessive global synchronisation to avoid causality problems. The performance analysis of \textit{optimistic} PDES for real-world large-scale network topology and complex QoS mechanisms is still inadequate. 
To address these gaps, this paper proposes a simulation toolkit, \textsc{Quaint}, which leverages an optimistic PDES engine ROSS, for detailed modelling of DiffServ-based networks. A novel event-handling model for each network router is also proposed to significantly reduce the number of events in complex QoS modelling.
\textsc{Quaint} has been evaluated using a real-world metropolitan-scale network topology with 5,000 routers/switches.
Results show that compared to the conventional simulator OMNeT++/INET, even the sequential mode of Quaint can achieve a speedup of 53 times, and the distributed mode has a speedup of 232 times. 
Scalability characterisation is conducted to portray the efficiency of distributed execution and the results indicate the future direction for workload-aware model partitioning.
\keywords{Network Digital Twin  \and Parallel Discrete Event Simulation \and Differentiated Services \and ROSS \and Optimistic Synchronisation. }
\end{abstract}
\setcounter{footnote}{0} 

\section{Introduction}


The Digital Twin (DT) paradigm, referring to the modelling, monitoring and continuous analysis of real-world assets or systems, is increasingly applied to the operation of communication networks \cite{almasan_network_2022,jang_digital_2023}.
A DT can be viewed as a model of the real-world system. DT assimilates real-time data from the network to update the model states, and leverages simulative what-if analysis to inform online optimisation of the real-world network \cite{almasan_network_2022,thelen_comprehensive_2022,zhang_knowledge_2024}.
Network operators can use DT to evaluate and predict network performance (e.g. throughput or latency) under different configurations without affecting the real-world network. DTs can also enable automatic, resilient, and full life-cycle operation and maintenance for networks \cite{irtf-nmrg-network-digital-twin-arch-04}, such as automatic service self-healing \cite{yu_digital_2023}. 

One of the critical challenges in building a DT is the high-fidelity modelling of the real-world network to support online predictions and what-if simulation analysis based on the real-time state of the network \cite{thelen_comprehensive_2022,zhang_knowledge_2024}. 
Although deep learning-based data-driven modelling approaches exist, they suffer from difficulty in collecting abundant datasets from the real network, long training time, generalisation problems, and inadequate explainability of the prediction results \cite{hui_digital_2022,almasan_network_2022,ZXTX20230619004}. 
High-fidelity packet-level Discrete Event Simulation (DES) tools are needed to either generate training data for deep learning models, or be directly used as a DT for online simulation analysis due to their accuracy and explainability. When used for online simulation, the DES should be continuously updated with real-time data of the network state to make timely predictions.

Challenges exist in the current research of DES for networks to support the online simulation capability of the DT. 
Widely adopted packet-level DES toolkits such as OMNeT++\footnote{https://omnetpp.org/} and ns-3\footnote{https://www.nsnam.org/} suffer from long execution times \cite{hui_digital_2022}. 
Although researchers have used Parallel Discrete Event Simulation (PDES) for network analysis \cite{gao_dons_2023,bai_unison_2024,nikolaev_performance_2013}, many works are based on \textit{conservative} simulators, which can suffer from excessive global synchronisation \cite{carothers_on_deciding_2010}. 
Current PDES studies have largely focused on simplified or standard network topologies \cite{nikolaev_performance_2013,wolfe_modeling_2018}. Studies on the impact of Quality-of-Service (QoS) models on the efficiency of the simulator are still emerging \cite{mcglohon_exploration_2021}. 
QoS mechanisms are common in modern networks and can add extra complexity to the modelling and simulation.
Differentiated Services (DiffServ) is one of the significant architectures for QoS, which contains per-hop forwarding behaviours, packet classification functions, and traffic conditioning functions including metering, marking, shaping, and policing \cite{rfc2475}. Modelling a token-bucket-based shaper with state-of-the-art toolkit INET\footnote{INET is a network model library of OMNeT++. See https://inet.omnetpp.org/} requires excessive periodic events to generate tokens \cite{token_bucket_inet}, which significantly increases the execution time.

This paper strives to address the above challenges by proposing a simulation toolkit, \textsc{Quaint} (\textbf{QU}ality-of-service-\textbf{A}ware \textbf{I}ntelligent \textbf{N}etwork digital \textbf{T}win), which utilises \textit{optimistic} PDES for large-scale DiffServ-based networks. 
Optimistic simulation is able to achieve low synchronisation overhead and high scalability \cite{carothers_on_deciding_2010}.
In particular, the novel contributions of this paper are as follows:
\begin{itemize}
    \item One of the earliest studies utilising \textit{optimistic} PDES to analyse networks with full pipelines of DiffServ QoS mechanisms.
    \item A highly efficient event-based simulation model for DiffServ-aware routers that significantly reduces the number of simulation events. Experiments show that \textsc{Quaint} can achieve 53 times speedup even in sequential mode compared with OMNeT++/INET.
    \item Performance characterisation for the scalability of optimistic PDES for a DiffServ-aware real-world large-scale bearer network. Results show 232.88 times speedup compared with OMNeT++/INET. The characterisation also points out the need for workload-aware model partitioning algorithms.
    
\end{itemize}

The rest of the paper is organised as follows: Section \ref{sec:background} provides the background on PDES and DiffServ. Section \ref{sec:related} reviews the related work in PDES for QoS-aware networks. Section \ref{sec:modelling} provides a novel router model. The performance characterisation of \textsc{Quaint} is presented in Section \ref{sec:evaluation}. Section \ref{sec:conclu} concludes the paper.

\section{Background} \label{sec:background}

\subsection{Parallel Discrete Event Simulation (PDES)}
In DES, the behaviour of a network is modelled as a sequence of events happening at specific points in time, such as packet arrival and sending. The simulator maintains an event queue and sequentially executes the events in timestamp order.
PDES method divides the network into several sub-networks, and each sub-network is simulated through a separate simulation process called a Logical Process (LP). Each LP is executed independently and can be viewed as a sequential simulator with its own event queue and local clock. LPs send time-stamped messages to schedule events for each other \cite{fujimoto_network_2007}.

LPs need to be synchronised to avoid causality errors, such that all events in each LP are processed in timestamp order. Causality errors occur when an LP receives events from other LPs with timestamps earlier than the LP's local clock.
There are two synchronisation approaches: conservative and optimistic.
Conservative synchronisation ensures that LPs do not receive any events with earlier timestamps, which is achieved via global synchronisation among LPs and the use of the lookahead value (the ability to predict the timestamp of the next event over a link) \cite{fujimoto_network_2007}. 
In general, the conservative approach performs well in systems with good lookahead properties.
In contrast, optimistic synchronisation does not force events to arrive in strict timestamp order, owing to its ability to detect and remedy causality violations of events. Each LP runs independently and only if an LP receives an event with an earlier timestamp than an event it has already processed, the LP will roll back to the previous state and re-process the events in timestamp order. The optimistic method can avoid unnecessary LP blocking and is less dependent on information about specific application scenarios, such as network delay, thus providing transparency to the applications \cite{fujimoto_network_2007}.   
The optimistic approach performs well in systems with a well-balanced workload, while it usually requires more memory because of the need to store the history of previous states to perform rollbacks \cite{fujimoto_network_2007}.

\subsection{DiffServ}
DiffServ is a widely adopted solution to scalable QoS mechanisms for IP networks \cite{curry_diffserv_pdes_2003,rfc2475}.
It differentiates incoming network traffic into different service classes, and applies differentiated traffic management policies. 
At the boundary nodes of a network, traffic is classified, conditioned, and assigned to different behaviour aggregates \cite{rfc2475}. Each aggregate is managed by different traffic policies at each node that allocate resources (e.g. buffer and bandwidth) according to per-hop behaviour (PHB) descriptions. 

DiffServ may involve several traffic conditioning elements: \textit{meter}, \textit{marker}, \textit{shaper}, and \textit{dropper}. Not all of the four elements are necessarily present in the conditioning process.
A meter measures the rate and burst size of the incoming traffic. A meter passes state information to other conditioning functions to trigger different conditioning behaviours.
A marker modifies the IP header of a packet and assigns it to a specific behaviour aggregate.
A shaper usually has a buffer and delays the packets to comply with a traffic profile, such as limiting the traffic forwarding rate.
A dropper selectively discards packets, thus achieving ``policing'' of packet streams \cite{rfc2475}.
The token bucket model is commonly applied to metering, shaping and policing. Tokens are generated continuously according to a predefined rate to a bucket with finite capacity. An incoming packet must consume a certain amount of tokens in the bucket to get through. If the tokens in the bucket are not enough, the packet shall wait, be discarded, or be marked. One common token-bucket-based model is the Single Rate Three Meter Marker (srTCM), which utilises two token buckets to meter the traffic rate and mark the packet as either green, yellow, or red \cite{rfc2697_srtcm}.

\section{Related Work} \label{sec:related}

This section reviews some of the most relevant studies about applying PDES for QoS-aware large-scale networks and analyses the differences with this paper.

Several recent studies have proposed optimised designs of \textit{conservative} PDES toolkits for networks to improve execution efficiency.
DONS \cite{gao_dons_2023} adopts Data-Oriented-Design to optimise the memory caching, which involves a complete re-architecting of the existing simulator design. Some selected QoS-related mechanisms are supported by DONS, such as TCP congestion control, Random Early Detection (RED) \cite{Floyd_RED_1993} and queue schedulers.
Unison \cite{bai_unison_2024} is another tool that achieves parallel-efficient and user-transparent network simulation.
The evaluation of Unison reports to incorporate selected QoS mechanisms for small-scale networks with no more than 50 nodes.
Some earlier work evaluates the distributed version of ns-3 for a billion network nodes, but the simulated network is simplified with single-hop traffic only and no QoS mechanisms.
\cite{nikolaev_pushing_2015,nikolaev_performance_2013}.
INET provides a versatile library for implementing QoS mechanisms. However, to execute INET in the parallel and distributed mode, a significant amount of implementation effort is needed to manually set up its engine OMNeT++,
as mentioned in \cite{gao_dons_2023}. This paper instead eliminates the daunting configuration effort by leveraging a PDES engine, which is natively designed for parallel and distributed execution.

In terms of \textit{optimistic} PDES, ROSS is one of the very few existing solutions and has achieved high efficiency and scalability \cite{carothers_ross_2002}. Built on ROSS, the CODES toolkit provides the model for exascale HPC (High-Performance Computing) networks \cite{mubarak_codes_2017}. 
Some work has used CODES to analyse flow control mechanisms for congestion avoidance \cite{wolfe_modeling_2018,mcglohon_exploration_2021,mubarak_evaluating_2019,brown_tunable_2021}. In \cite{wolfe_modeling_2018}, a credit-based mechanism is proposed for slim-fly HPC networks, which allows the receiver router to notify the sender about the buffer availability by sending a credit.
In \cite{mcglohon_exploration_2021}, the authors have further studied a three-phase congestion control approach for Dragonfly HPC networks. The approach detects congestion by monitoring the output buffer usage, identifies the cause and then abates the congestion by sending a signal back to the source.
Some other works have used CODES to evaluate the network performance of QoS mechanisms that differentiate traffic classes. 
The work in \cite{mubarak_evaluating_2019} models the behaviour of bandwidth monitoring and capping, and evaluates the network performance of two traﬃc prioritisation mechanism configurations.
Prioritised packet sending and a shaping strategy similar to Two Rate Three Color Marking (trTCM) for HPC networks has been proposed in \cite{brown_tunable_2021}.  

However, neither DONS nor Unison focus on detailed and complex QoS modelling and performance evaluation. 
CODES mainly focuses on HPC networks, whose workload and network topology characteristics can largely differ from city-scale networks, e.g., HPC network nodes have lower link latency and simplified QoS mechanisms \cite{mubarak_evaluating_2019}.
Although CODES supports several QoS models, these models are simplified and non-standard, partly because QoS is not widely studied in HPC networks compared to TCP/IP networks \cite{mubarak_evaluating_2019}. 
Each CODES-related work focuses on only one or a few QoS components, and the composite effect with multiple QoS modules on the simulator's performance is unknown.

Table~\ref{tab1} gives a summary of the relevant PDES toolkits. This paper proposes \textsc{Quaint}, which utilises optimistic scheduling and is evaluated with fine-grained DiffServ modules and a non-standard city-scale network topology. 

\begin{table}
\caption{Comparison between the state-of-the-art parallel discrete event network simulation toolkits for QoS analysis.}\label{tab1}
\begin{tabular}{|m{5.5em}|m{19em}|l|m{4.8em}|m{5em}|}
\hline
Toolkit & QoS Modelling & Sync\textsuperscript{a} & Network Topology\textsuperscript{b} & Simulator Kernel\\
\hline
\hline
DONS \cite{gao_dons_2023} &  TCP Congestion control, RED, schedulers & C & Standard and non-standard & Self-developed using Unity \\
\hline
Unison \cite{bai_unison_2024} & TCP Congestion control, RED  & C & Standard & Modified ns-3 \\
\hline
\multirow{4}{5.5em}{CODES} & Congestion control \cite{wolfe_modeling_2018,mcglohon_exploration_2021} & \multirow{4}{1em}{O} & \multirow{4}{4.8em}{Standard HPC} & \multirow{4}{5em}{ROSS} \\
& Bandwidth monitoring and capping \cite{mubarak_evaluating_2019}  &  & &  \\
& Prioritisation on differentiated classes \cite{mubarak_evaluating_2019,brown_tunable_2021}  &  & &  \\
& trTCM traffic shaping \cite{brown_tunable_2021} &  & &  \\
\hline
\textsc{Quaint} \ \ (this paper) & \textbf{DiffServ}: fine-grained modules for traffic classification and conditioning & \textbf{O} & \textbf{Non-standard} & ROSS \\
\hline
\end{tabular}

\vspace{1ex}
{\raggedright \textsuperscript{a} C for conservative, O for optimistic.\par}
{\raggedright \textsuperscript{b} ``Standard'' refers to topologies defined by equations, such as FatTree and Torus.\par}
\end{table}


\section{QoS Modelling in PDES} \label{sec:modelling}

To address the gap in PDES for fine-grained QoS modelling, this section proposes a model of fine-grained QoS mechanisms in DiffServ-aware networks. The model also aims to reduce unnecessary events to improve the simulation efficiency.

\subsection{A Motivating Example of DiffServ Pipelines}

\begin{figure}
\includegraphics[width=\textwidth]{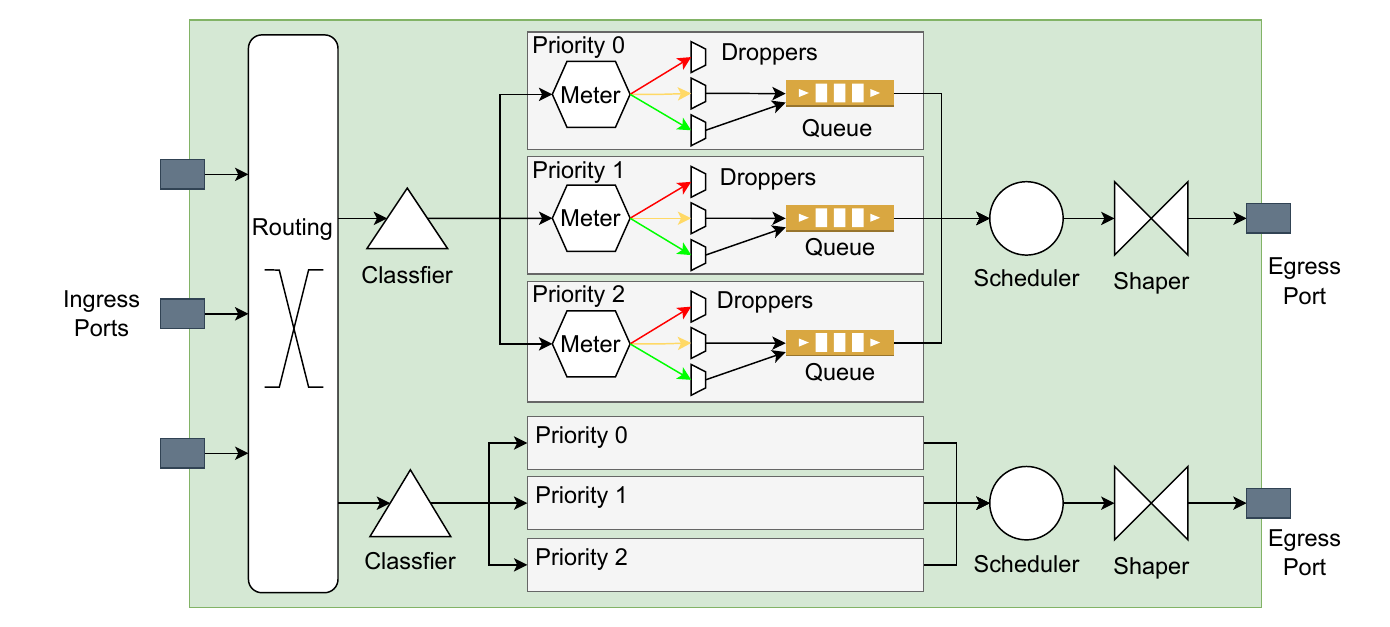}
\caption{The internal structure of a router under study. 
The number of ingress and egress ports can vary in different routers.} \label{fig-qos}
\end{figure}

The DiffServ framework adopted in the paper is shown in Fig.~\ref{fig-qos}, which shows a representative pipeline of QoS mechanisms within a single router. It captures all essential elements specified in DiffServ classification and conditioning for every egress port: classifier, meter, marker, dropper, and shaper.
For a packet outside the router, it first arrives through one of the ingress ports. Then according to the routing table, the packet is assigned to the QoS pipeline of a specific egress port. The corresponding classifier of the pipeline classifies the packet according to its DS field in the IP packet header to select a PHB \cite{rfc2475}. This paper implements different PHBs using a combination of meter, dropper, queue, scheduler and shaper with different parameters to achieve prioritisation. 
The classified packet is then directed to one of the three priority classes. An srTCM module \cite{rfc2697_srtcm} acts as both a meter and a marker to measure the packet rate (metering) and marks the packet with a color (marking). A dropper then decides whether to drop the packet based on its color. Packets with different colors will be managed by different droppers with specialised parameters. If the packet is not dropped, it is put in a queue to wait to be sent out.
A scheduler is attached to each egress port and uses specific policies to decide which queue to take out a packet. At the end of the pipeline, a token-bucket-based shaper limits the forwarding speed of the traffic and delays packets that exceed the permitted burst rate.

\subsection{Problems in Modelling Token Buckets}

The problem in modelling the QoS pipeline shown in Fig.~\ref{fig-qos} is in the token-bucket-based shaper. 
A token bucket can be viewed as a container with a capacity $C$. New tokens are continuously generated into the bucket at the rate of $r$ per second. Overflowed tokens are discarded. Equivalently, if tokens are generated every $t$ second, then each time $(r\cdot t)$ tokens need to be added into the bucket. When a packet of $n$ bytes arrives, the packet needs to remove $n$ tokens from the bucket in order to be sent out. If there are not enough tokens in the bucket, the packet needs to wait until the bucket has accumulated to no less than $n$ tokens.

In the event-driven simulation, modelling token-bucket-based shaper requires repeatedly checking the availability of the bucket and generating new tokens. The state-of-the-art QoS modelling toolkit INET of OMNeT++ \cite{token_bucket_inet} shows that the token bucket module requires a fixed interval of token generation events, thus significantly increasing the total number of events and reducing the simulation execution speed, causing performance bottleneck.

Although token buckets are also used in srTCM, the input packet processing will not be delayed as in shapers. Packets will always be immediately passed into the next dropper module without re-processing the same packet in the future. While a shaper relies on its own state (number of tokens) and clock to decide when to handle a packet again if the state not permitting.

The subsequent subsection presents a solution to the aforementioned problem of excessive periodic token generation events, which employs a flag-based event handling mechanism to eliminate the need for token generation at fixed time intervals.




\subsection{Event Handling Model}

\begin{figure}
\includegraphics[width=\textwidth]{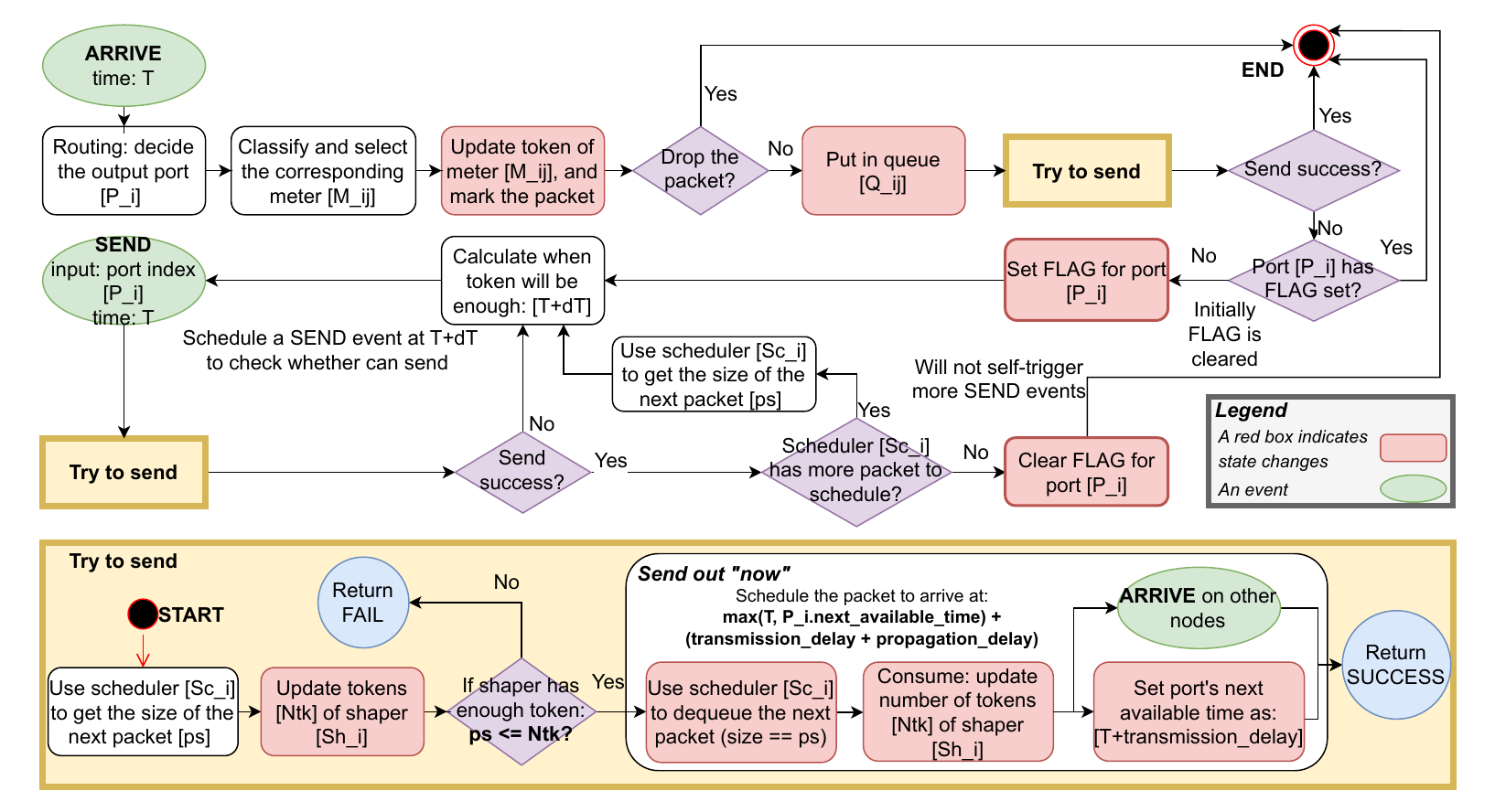}
\caption{The proposed event handling model of a router. The workflow starts from an event (a green ellipse) and ends at a sink or creating a new event. A ``try to send'' procedure is extracted from the main workflow and shown at the bottom.} \label{fig-model}
\end{figure}

Considering the characteristics of token-bucket-based shapers, this paper abstracts the most essential modules of DiffServ and proposes a high-efficiency event-handling model of a DiffServ-aware router. The overall workflow of the model is shown in Fig.~\ref{fig-model}.
To avoid complex event handling logic, only two types of events are involved: \verb|ARRIVE| and \verb|SEND|.
The basic idea is to only schedule events at the virtual timestamps when a packet arrives and when the router is predicted to accumulate enough tokens to forward a pending packet, rather than repeatedly scheduling token generation events at a fixed time interval as in INET.  
When a packet arrives at a router, it triggers an \verb|ARRIVE| event. All QoS modules are activated and updated during the handling of the \verb|ARRIVE| event. 
If the shaper does not have enough tokens, the packet is queued, and the \verb|ARRIVE| event will trigger a future \verb|SEND| event to try to send later when tokens are enough.
Each port has a try-to-send pipeline. The \verb|SEND| is only a trigger for performing a try-to-send behaviour of a port, but is not tied to any specific packet.
However, if too many packets are queued, they can all trigger \verb|SEND| events in the future, which can be redundant if a \verb|SEND| can send out multiple packets.
A proper design is needed since determining the timestamp of a future \verb|SEND| requires the awareness of both the token generation time $\Delta t$ for previous pending packets and the scenario where higher-priority packets arrive earlier before now + $\Delta t$.
To solve this problem, the model draws inspiration from CODES\footnote{https://github.com/codes-org/codes/blob/b8c423e3426f2793089de5b43b2da2086\\9eb2f77/src/networks/model-net/dragonfly-dally.C} and uses a flag for each egress port.
If the QoS pipeline of an egress port is idle (not forwarding any packet), the first \verb|ARRIVE| event must trigger the try-to-send function. This is done to respect the order and precedence of packets. The flag prevents future \verb|ARRIVE| events from triggering a redundant try-to-send. The first \verb|ARRIVE| event sets this flag and stays active until this egress port becomes idle. 
This design will significantly reduce redundant events, thus increasing execution speed (see Section \ref{sec:evaluation}).
This model also handles the scenario where packets differ in size and priority.





\begin{figure}[t]
\centering
\includegraphics[width=0.6\textwidth]{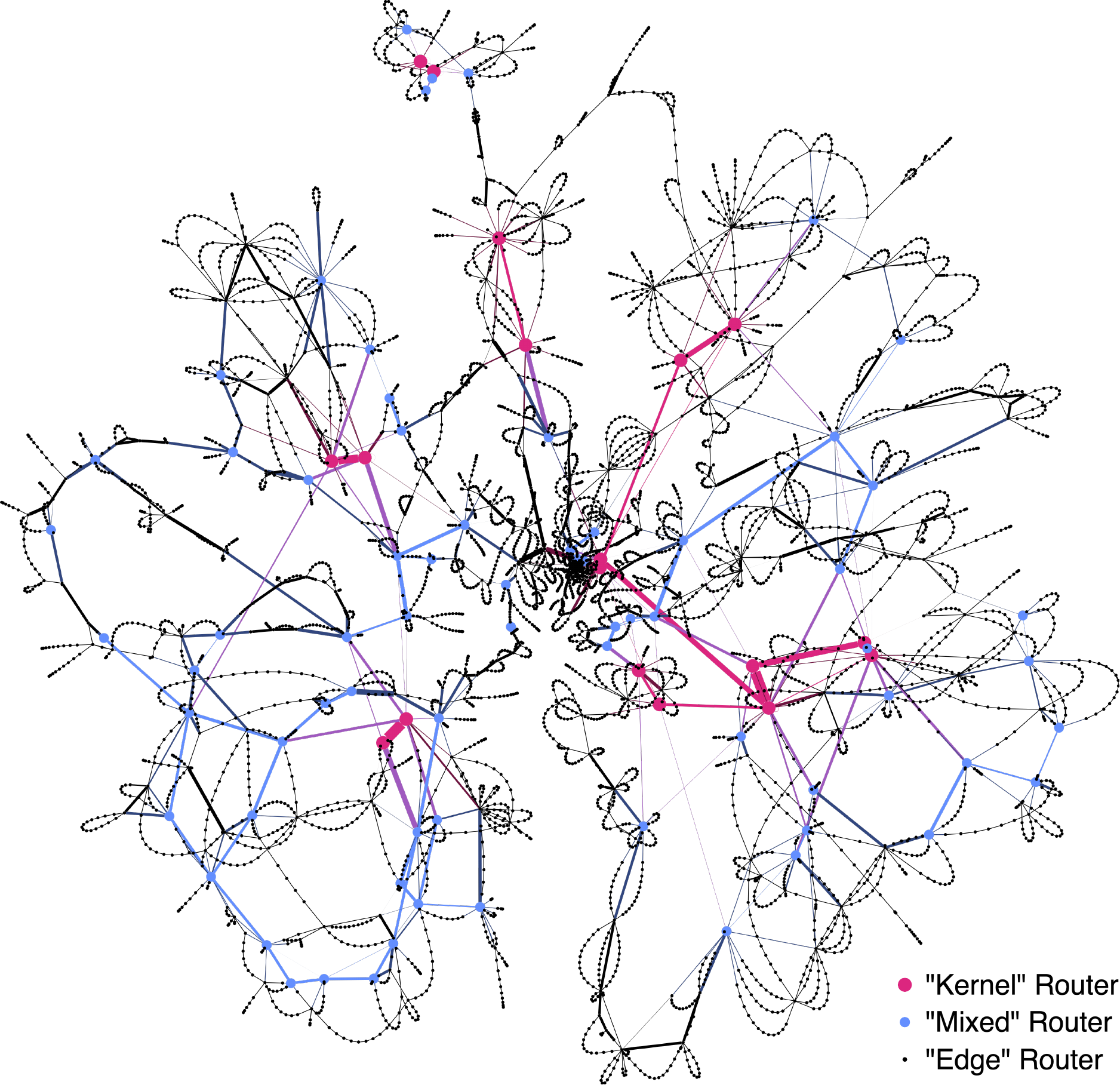}
\caption{Topology of a real-world metropolitan-scale bearer network. Black dots are ``access'' routers. Blue dots are ``mixed'' routers. Red dots are ``kernel'' routers. An access router usually has 1 to 3 ports with 25Gbps bandwidth. A mixed or access switch can have over 10 ports with 10-100Gbps bandwidth. } \label{fig-network}
\end{figure}

\section{Evaluation with Real-World Topology} \label{sec:evaluation}

To evaluate the performance of the proposed simulation model with optimistic parallel and distributed execution, this paper developed \textsc{Quaint}, a network simulator based on the ROSS\footnote{https://ross-org.github.io/} simulation engine. The source code, data and parameters used for all the experiments are publicly available online\footnote{Source code, data and parameters: https://github.com/network-digital-twin}.
The evaluation objective is to examine the accuracy and execution efficiency of \textsc{Quaint} when simulating real-world large-scale network topologies, specifically: 1) to assess the execution time of the proposed simulator compared with OMNeT++, which is a widely-adopted network simulator and has a versatile model library, INET, for comprehensive QoS modelling; and 2) to investigate the scalability of the proposed simulator with different numbers of processors.

The benchmark model is built according to the topology data of a real-world bearer network, shown in Fig.~\ref{fig-network}. A bearer network connects base stations and data centers, with a large number of nodes\footnote{This paper refers to a ``node'' as either a router or a switch.} deployed across cities \cite{jiang_zte_bearer}. Fig.~\ref{fig-network} contains 5,237 nodes and 6067 links, with 5149 \textit{access} nodes, 70 \textit{mixed} nodes, and 18 \textit{kernel} nodes.
Each node is configured with the shortest path routing.
All nodes are modelled according to Fig.~\ref{fig-qos}, with srTCM-based metering and marking, RED droppers \cite{rfc2309,Floyd_RED_1993}, strict-priority schedulers, and token-bucket-based shapers. 

All experiments in this paper were run on a cluster of four DELL PowerEdge T64 servers. Each server has 40 CPU cores (Gold 6230 2.1G*2), 256G RAM, and 1.92TB SSD. All servers are connected via 56Gbps InfiniBand FDR to a Mellanox SX6036 switch. Each server is installed with Ubuntu 20.04.6 LTS, OpenMPI 4.1.6 and UCX 1.15.0.

\subsection{Comparison with OMNeT++/INET}

\begin{figure}
\centering
    \begin{subfigure}[b]{0.23\textwidth}
        \centering
        \includegraphics[width=\textwidth]{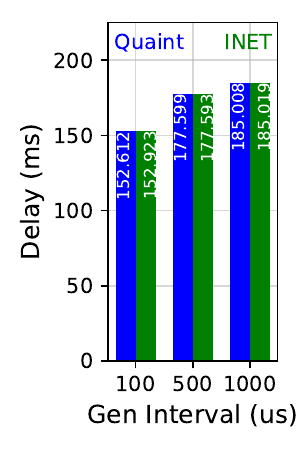}
        \caption{Delay}
        \label{fig:sub-delay}
    \end{subfigure}
    \hfill
    \begin{subfigure}[b]{0.23\textwidth}
        \centering
        \includegraphics[width=\textwidth]{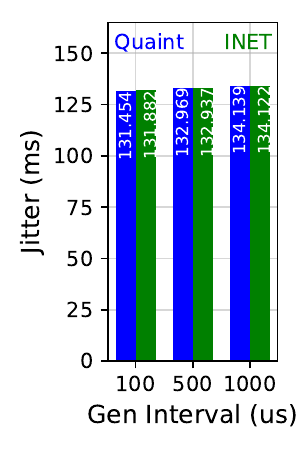}
        \caption{Jitter}
        \label{fig:sub-jitter}
    \end{subfigure}
    \hfill  
    \begin{subfigure}[b]{0.23\textwidth}
        \centering
        \includegraphics[width=\textwidth]{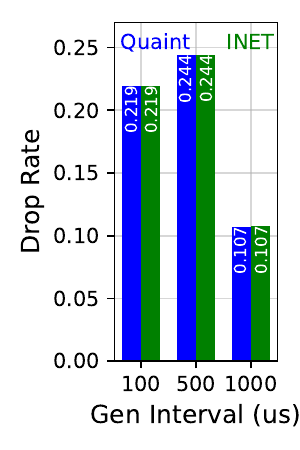}
        \caption{Drop rate}
        \label{fig:sub-drop}
    \end{subfigure}
    \hfill
    \begin{subfigure}[b]{0.23\textwidth}
        \centering
        \includegraphics[width=\textwidth]{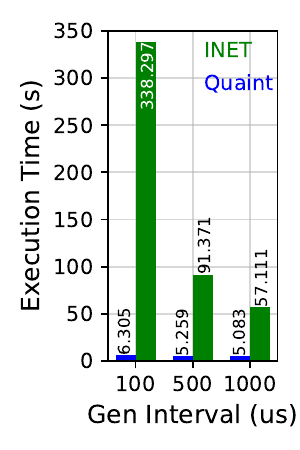}
        \caption{Runtime}
        \label{fig:sub-execution}
    \end{subfigure}
    \caption{Comparison of the accuracy and efficiency between \textsc{Quaint} and INET}
    \label{fig:inet-ross}
\end{figure}

In order to evaluate the efficiency and accuracy of the proposed simulation model in Section \ref{sec:modelling}, the proposed simulator is compared with INET.
The evaluation uses synthetic traffic in which each \textit{access} router continuously sends packets to one random \textit{mixed} or \textit{kernel} router. The simulation lasted for 500,000,000 ns virtual time, and around $5\times 10^5$ packets (each packet is 1400 bytes) were simulated. 
In this experiment, \textsc{Quaint} only runs in sequential mode on one server, i.e. only one process is used to run the simulation, and no parallelisation is involved.  Since INET uses a fixed token generation interval for the token bucket-based shaper, the two simulators were compared under varying values of the token generation interval. Each configuration was repeated 5 times. The results are shown in Fig.~\ref{fig:inet-ross}.

As can be seen from  Fig.~\ref{fig:sub-delay} \ref{fig:sub-jitter} and \ref{fig:sub-drop}, the simulated network performance (average delay, jitter and drop rate of all packets) of \textsc{Quaint} has extremely small differences compared to that of INET.
\textsc{Quaint} also largely reduces the execution time due to the special modelling of token generation, as shown in Fig.~\ref{fig:sub-execution}.
It should be noted that in order to mirror the behaviour of a real-world router accurately, the generation interval in the model should be small enough. 
For instance, a 15Gbps ingress port can witness packets (1400B each) arriving at most every 0.747 $\mu s$. A generation interval multiple magnitudes larger than 0.747, e.g. 1000 $\mu s$, may increase the packet delay, thus deviating from the behaviour of the reality, which is shown in Fig.~\ref{fig:sub-delay}.
With the token generation interval as small as 100 $\mu s$, \textsc{Quaint} running in sequential mode (no parallelisation) can execute 53.66 times faster than INET.



\subsection{Scalability}
Then \textsc{Quaint} was evaluated for its scalability against varying numbers of processes.
The traffic workload follows the same generation logic as the previous section, but the simulation time is extended to $10^9$ ns. 
In order to partition the network into different processes, the graph partitioning tool METIS was used \cite{metis}.
Processes are evenly distributed onto the servers and we ensure each 40-core server runs no more than 20 processes while minimising the number of servers used. In particular, experiments with less than or equal to 20 processes were run on one server.

\begin{figure}[b]
\centering
    \begin{subfigure}[b]{0.23\textwidth}
        \centering
        \includegraphics[width=\textwidth]{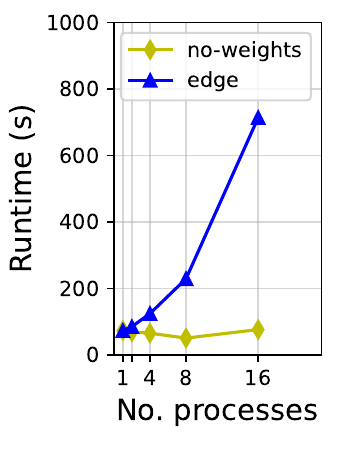}
        \caption{Edge weights}
        \label{fig:sub-edge}
    \end{subfigure}
    \hfill
    \begin{subfigure}[b]{0.35\textwidth}
        \centering
        \includegraphics[width=\textwidth]{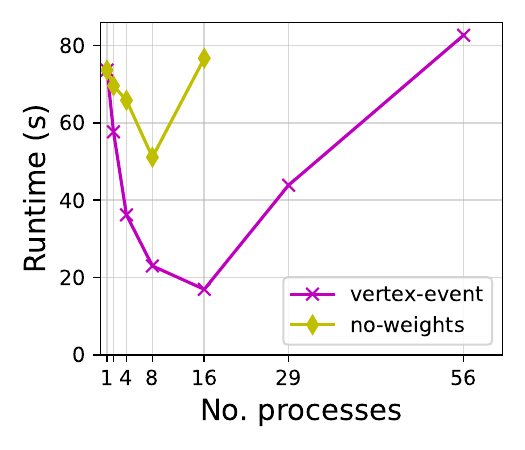}
        \caption{Cut-off point improvement}
        \label{fig:sub-vertex-evt}
    \end{subfigure}
    \hfill
    \begin{subfigure}[b]{0.35\textwidth}
        \centering
        \includegraphics[width=\textwidth]{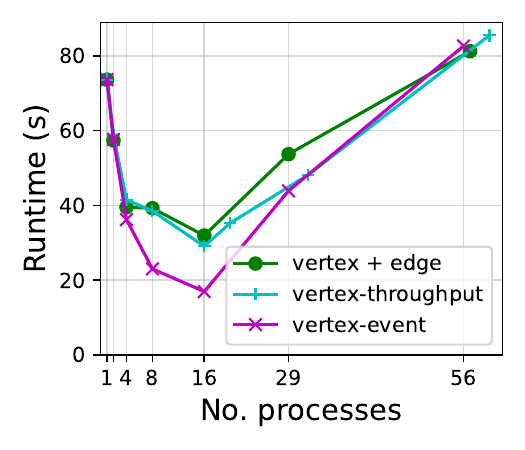}
        \caption{Weight comparison}
        \label{fig:sub-vertex-pkt}
    \end{subfigure}
    \caption{The execution time of the proposed simulator given different levels of parallelisation (number of processes) using different partitioning approaches.}
\label{fig:scalability}
\end{figure}

\begin{figure}[t]
\centering
    \begin{subfigure}[b]{0.45\textwidth}
        \centering
        \includegraphics[width=\textwidth]{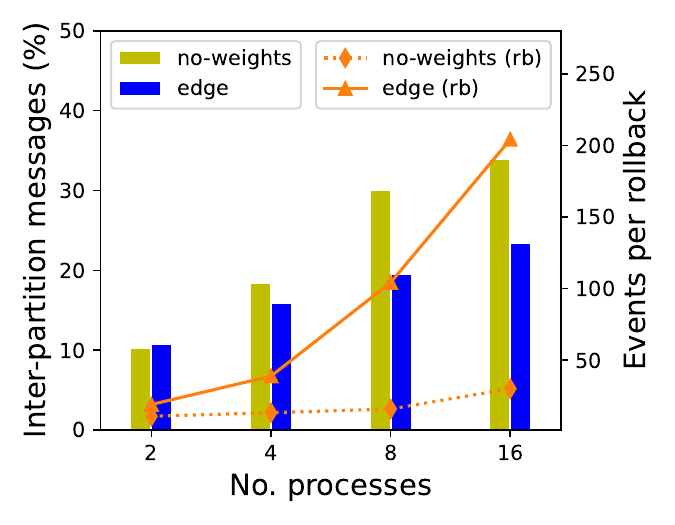}
        \caption{}
        \label{fig:sub-edge-rb}
    \end{subfigure}
    \hfill
    \begin{subfigure}[b]{0.412\textwidth}
        \centering
        \includegraphics[width=\textwidth]{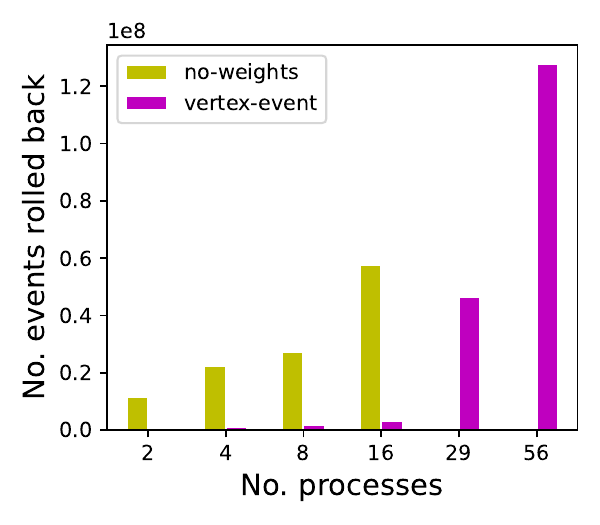}
        \caption{}
        \label{fig:sub-vertex-rb}
    \end{subfigure}
    \caption{Characterisation on rollbacks and inter-partition messages. The result of \textit{no-weights} for more than 16 processes is not included.}
\label{}
\end{figure}

An intuitive idea for partitioning the network model is to minimise the exchange of inter-partition messages/packets. However, for optimistic simulation, this approach turned out to be highly inefficient. 
We assigned the throughput (calculated by packets per second, since each packet has the same size) of each edge as the edge weight, and used METIS to minimise the sum of the weights of all straddling edges. 
The results in Fig.~\ref{fig:sub-edge} (\textit{edge}) show that more processes will only increase the execution time, and even worse than not assigning weights at all (\textit{no-weights}).
As shown in Fig.~\ref{fig:sub-edge-rb}, although the number of inter-partition packets (bars in the figure) of \textit{edge} are indeed reduced, each rollback message will trigger more events (lines in the figure) to be rolled back than \textit{no-weights}, which indicates that each partition progresses in virtual time with different speed. Messages sent from a straggler partition with a slower clock may cause other partitions to roll back much farther back in virtual time, thus increasing the number of events to be re-processed.
Therefore, the key to partitioning in optimistic network simulation is to balance the workload in each partition such that their clock progresses at similar speed at wall-clock time to avoid the straggler effect.

One solution identified by this paper is to properly assign weights to vertices (the network nodes): the weight of each vertex is assigned to be the total number of events to be processed by that network node, which indicates the computation workload intensity. 
The idea is to ensure an even distribution of workload among various partitions.
The evaluation result is shown as \textit{vertex-event} in Fig.~\ref{fig:sub-vertex-evt}. 
The figure shows continuous speedup until the cut-off point of 16 processes, with 4.34 times speedup. While as the number of processes increases further, the system becomes communication-bound: the overhead of inter-partition communication and rollbacks cancels the benefit of parallelisation. As shown in Fig.~\ref{fig:sub-vertex-rb}, the number of events rolled back (rollback overhead) increases dramatically after more than 16 processes.

The number of events may not be easily known in advance in practice, but can be estimated by the ingress throughput of the network node: more arrival packets relate to more events. The result of this approach is shown as \textit{vertex-throughput} in Fig.~\ref{fig:sub-vertex-pkt}, which has a similar trend and cut-off point compared to \textit{vertex-event}. The configuration \textit{vertex + edge} further integrates \textit{vertex-throughput} with edge weights defined in \textit{edge},  but shows an overall slowdown near the cut-off point compared to \textit{vertex-throughput} alone.
This coincides with the observation in Fig.~\ref{fig:sub-edge} that partitioning based on minimising the total weights of straddling edges does not contribute to more efficient execution in optimistic simulation.


\subsection{Summary and Discussion}

The evaluation of sequential execution compared with OMNeT++/INET has shown the significant speedup achieved by \textsc{Quaint}. INET's design trades performance for usability, leaving space for the proposed simulator \textsc{Quaint} to improve the performance bottleneck. 
The approaches of existing related studies rely on heavily modifying the existing kernel or redesigning a new kernel. This paper instead uses a lightweight versatile PDES kernel and adapts it to be a network simulator, allowing users to leverage the benefit of PDES's APIs without heavily redesigning the simulator and the scheduling mechanisms.

The scalability evaluation demonstrates the great potential of the distributed version of \textsc{Quaint}. The analysis also shows the need for proper model partitioning, which significantly affects the simulator's efficiency. With a proper design of the partitioning, the cut-off point can be largely improved.
The characterisation indicates the need to design partitioning algorithms that balance the computation workload in different partitions. 





\section{Conclusion}\label{sec:conclu}

This paper has proposed \textsc{Quaint}, an optimistic PDES toolkit for QoS-aware networks, including fine-grained DiffServ models and a highly efficient event-handling workflow model. \textsc{Quaint} has been evaluated using the topology of a real-world large-scale network. According to the evaluation, the distributed version of \textsc{Quaint} has demonstrated 4.34$\times$53.66=232.88 times speedup compared to OMNeT++/INET. The performance characterisation suggests further research in partitioning algorithms to balance computation workloads across the partitions. Future work will also evaluate different traffic workloads and QoS mechanisms.
This work provides pioneering indications for utilising optimistic PDES for modelling complex QoS mechanisms and performance analysis.


\subsubsection*{Acknowledgements.}
This work is supported by: ZTE Communication Technology Service Co., Ltd., China (Project No: IA20230803007); SUSTech Research Institute for Trustworthy Autonomous Systems, China.
%
%
%
\bibliographystyle{splncs04}
\bibliography{mybibliography}

\end{document}